\documentclass[10pt, aps,prc,twocolumn,superscriptaddress,preprintnumbers,
amsmath, 
floatfix,
longbibliography,
nofootinbib
]{revtex4-1}
\usepackage[T1]{fontenc}
\usepackage[utf8x]{inputenc} 
\usepackage{adjustbox}          
\usepackage[caption=false]{subfig}
\usepackage{url}
\usepackage{color}
\usepackage{float}
\usepackage[pdftex,colorlinks=true, linkcolor = blue, citecolor=blue,urlcolor=blue, bookmarksnumbered=true, bookmarksopen=true]{hyperref}
\usepackage{longtable}
\usepackage{amsfonts}
\usepackage{wrapfig,bm,bbm}
\newcommand{\beq}{\begin{equation}}
\newcommand{\eeq}{\end{equation}}
\newcommand{\bea}{\begin{eqnarray}}
\newcommand{\eea}{\end{eqnarray}}

\begin{document}

\title{Fission Fragment Excitation Energy Sharing Beyond Scission}
  
\author{Aurel Bulgac}%
\email{bulgac@uw.edu}%
\affiliation{Department of Physics,%
  University of Washington, Seattle, Washington 98195--1560, USA}
  
\date{\today}

\begin{abstract}
A simplified, though realistic, model describing two receding and accelerating fission fragments, 
due to their  mutual Coulomb repulsion, shows that fission fragments share excitation energy
well after they ceased to exchange nucleons. This mechanism leads to a lower total kinetic 
energy of the fission fragments, particularly if the pygmy resonances in the fission fragments are excited.  
Even though the emphasis here is on fission, similar arguments 
apply to fragments in heavy-ion reactions.
\end{abstract}

\preprint{NT@UW-20-01}

\maketitle

\section{Introduction}

Since the discovery of fission in 1939~\cite{Hahn:1939,Meitner:1939,Bohr:1939} it was assumed that after scission 
the two fragments are accelerated by their Coulomb repulsion and the entire 
potential Coulomb energy between the fragments is converted into the total kinetic energy (TKE)
of the fission fragments (FFs). With the exception of a couple of small studies of which I am aware 
of~\cite{Mustafa:1971,Bertsch:2019q,Carjan:2018}, this assumption is treated as rather accurate and 
the magnitude of the TKE of the FFs was used as a signature of the scission shape of the fissioning
nucleus or to disentangle different fission modes~\cite{Brosa:1990,Wagemans:1991,Sierk:2017,Usang:2019}.  
However, because of the long range nature of the Coulomb interaction the
intrinsic excitation energy can be still exchanged between the receding FFs and the amount of the 
total excitation energy (TXE) and TKE can be affected.  In a different kind of study, 
\textcite{Bertsch:2019q} argues that the long range Coulomb 
interaction between deformed FFs can lead to their re-orientation and as a result it can affect 
their angular momentum content.

\begin{figure}[t]
\includegraphics[width=0.9\columnwidth]{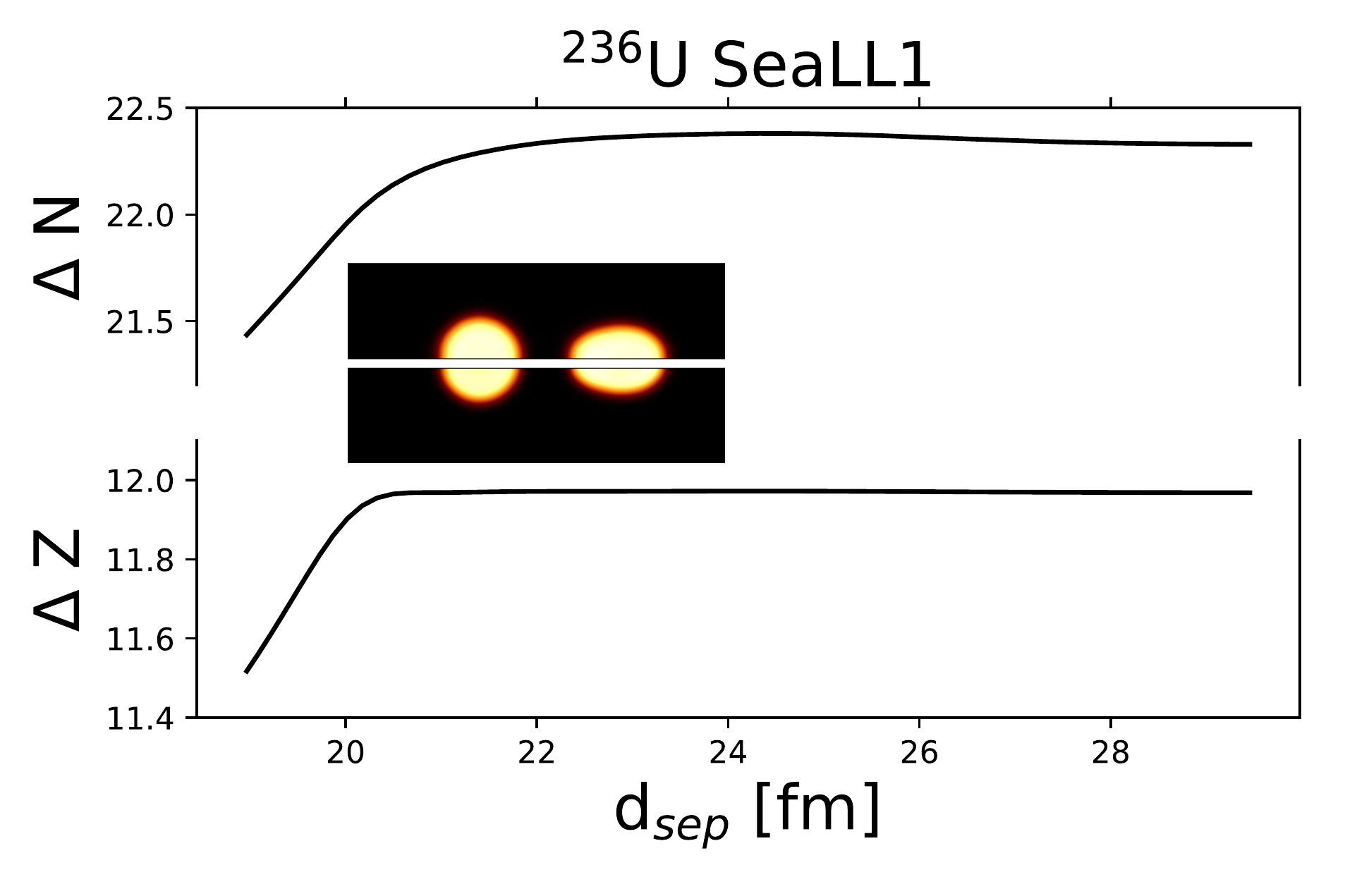}
\includegraphics[width=0.9\columnwidth]{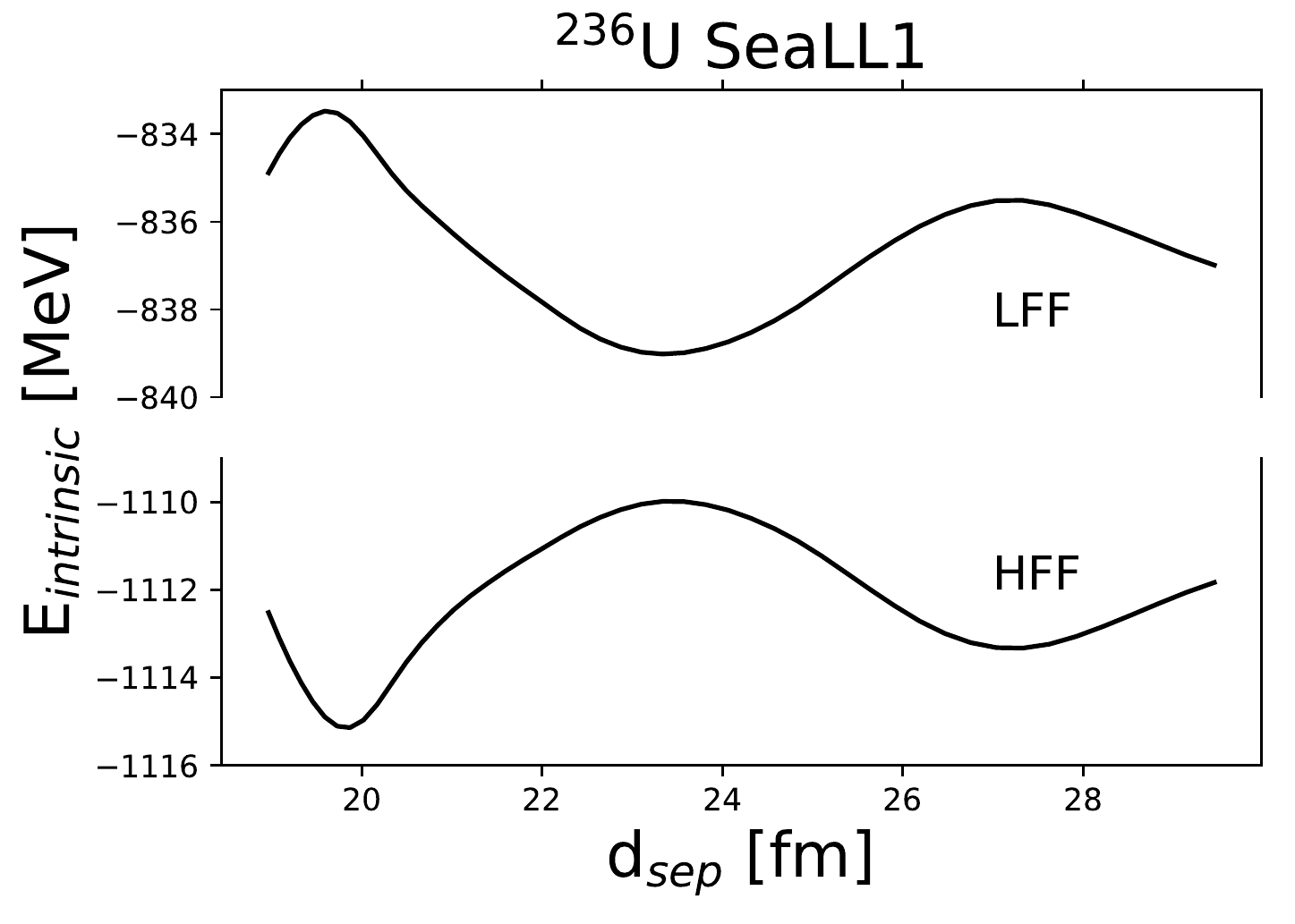}
\includegraphics[width=0.9\columnwidth]{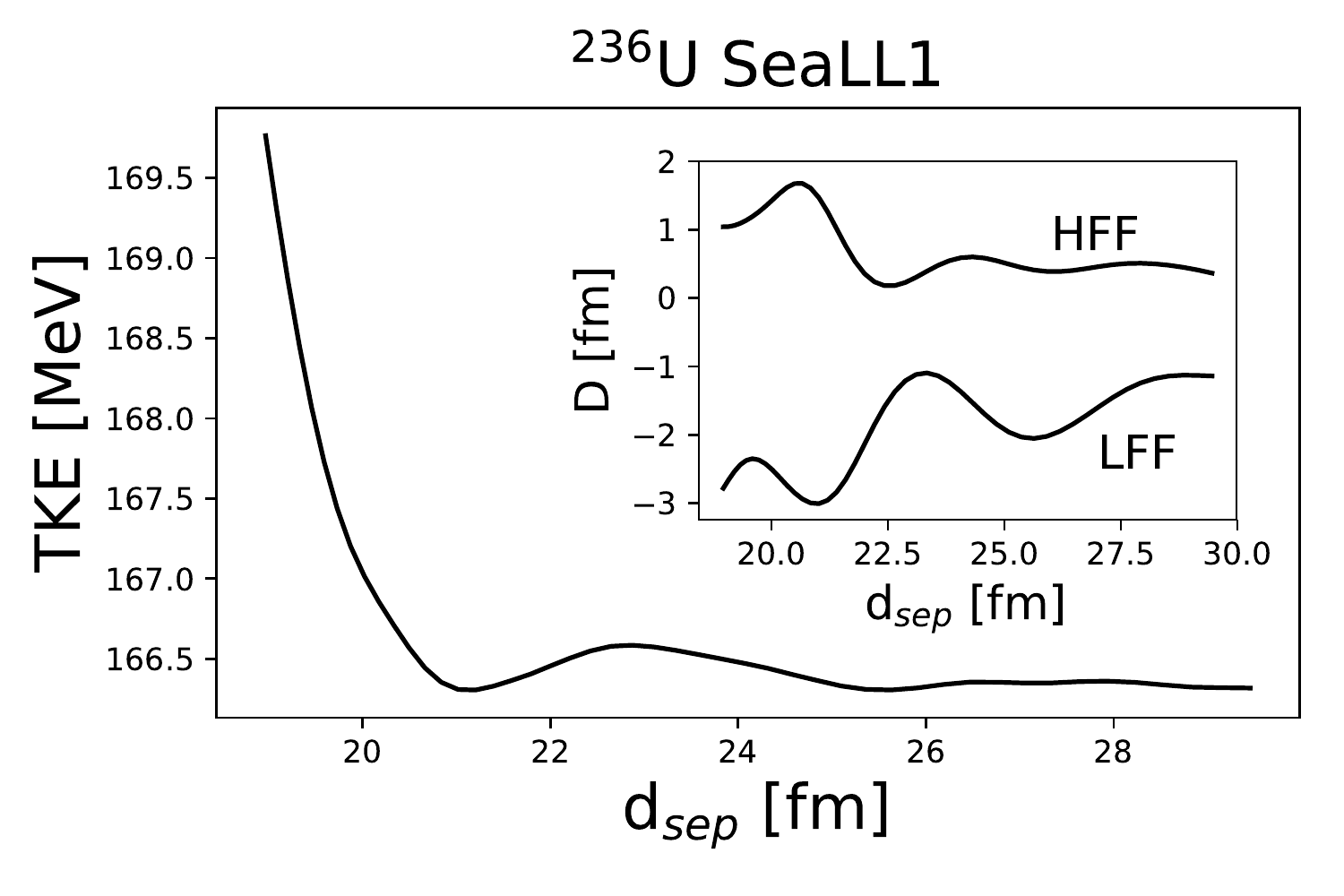}
\caption{ \label{fig:FFs} (Color online) The evolution of the neutron and proton number differences  
$\Delta N$ and $\Delta Z$ (upper panel), the FFs intrinsic energies (middle panel), and the
TKE and in the inset the FFs dipole moments $D$ (lower panel)
 as a function of the fission fragments separation d$_{sep}$ in the case of induced fission of $^{236}$U. 
 The numerical results are from a work in progress~\cite{IA:2020}. In the inset of the upper panel the 
 neutron (upper half)  and proton (lower half) shapes of FFs distributions are displayed at 22 fm separation. 
 The heavy FF is on the left and the light FF in on the right. Notice that the LFF is more elongated than the HHF. }
\end{figure}

In Fig.~\ref{fig:FFs} I illustrate these points using some typical results~\cite{IA:2020} obtained by simulating 
the induced fission of $^{236}$U  resulting from the reaction $^{235}$U(n,f),  
within the time-dependent density functional theory (TDTDF) framework
described in Refs.~\cite{Bulgac:2016,Bulgac:2019b,Bulgac:2020}, with the nuclear energy density functional (NEDF) 
SeaLL1~\cite{Shi:2018}. Similar results are obtained for other NEDFs. 
In the case illustrated in Fig.~\ref{fig:FFs} the scission occurs  when the separation between the FFs centers of mass d$_{sep}$
exceed about 21 fm. Before scission d$_{sep}$ is defined as the distance between the centers of mass 
of the two halves of the fissioning nucleus. The neck forms quite closely to the center of mass of the fissioning nucleus.  

At d$_{sep}$ = 21 fm separation there is practically no nucleon exchange 
between the FFs and the quantities
\begin{align} 
\Delta N=N_{HFF}-N_{LFF} , \quad \Delta Z=Z_{HFF}-Z_{LFF} 
\end{align}
attain
their asymptotic values, see the upper panel in Fig.~\ref{fig:FFs}. Here $N_{HFF,LFF}$ and $Z_{HFF,LFF}$ 
are the heavy and light FFs neutron and proton numbers respectively, calculated as the corresponding numbers of
of nucleons in the left and right halves of the simulation box. 
The center of mass of the system is exactly in the middle of the box. 
When the FFs are sufficiently well separated these 
are the actual FFs neutron and proton numbers.
Unlike protons, for which the Coulomb 
barrier hinders significantly their emission, a small neutron cloud is formed around the FFs, and the FFs neutrons numbers 
are not as sharply defined as the FFs proton numbers, see the upper panel in Fig.~\ref{fig:FFs}. 

Even though effectively no particle transfer occurs for d$_{sep}>  21$ fm 
between the FFs, when the separation between the tips of the two FFs exceeds the range of the nucleon interactions,
the FFs intrinsic energies change by significant amounts, with amplitudes of the order 
of several MeVs.
The middle panel of Fig.~\ref{fig:FFs} shows that the FFs intrinsic energies evolve 
as well and that they are out of phase, while their sum oscillates with a small amplitude, comparable to the TKE oscillation 
amplitude, shown in the lower panel of Fig.~\ref{fig:FFs}. 
This behavior of the FFs intrinsic excitation energies is in agreement with the Eqs.~(\ref{eq:xi1}-\ref{eq:xi2}) 
and the conservation of the sum of TKE and intrinsic energies $E_\text{tot} = E_\text{int}(t) + E_\text{TKE}(t)$, see Eq.~(\ref{eq:et}-\ref{eq:tke}).

In the inset of the lowest panel of Fig.~\ref{fig:FFs} I show the evolution of the FFs dipole moments, 
defined according to Eq.~(B.88) of Ref.~\cite{Ring:2004},
\begin{align}
D = \frac{NZ}{N+Z}(z_P-z_N),
\end{align}
where $z_{P,N}$ are the proton and neutron $z$ centers of mass coordinates ($Oz$ being the fission axis) 
and $N$, $Z$ the neutron and the proton numbers for each FF.
As one might have expected, the dipole moments of the two FFs  oscillate out of phase, see also Sec.~\ref{sec:II}, and since the LFF is 
more elongated along the fission $z$-axis than the HFF~\cite{Bulgac:2016,Bulgac:2019b,Bulgac:2020} its amplitude is larger. 
Clearly the two FFs exchange intrinsic energy due to the excitation of their respective lowest dipole modes, which 
in this case $\approx 6$ MeV.

 At each separation d$_{sep}$ (or corresponding time) the TKE is evaluated 
by adding together the FFs instantaneous kinetic and Coulomb FFs interaction energies~\cite{Bulgac:2016,Bulgac:2019b,Bulgac:2020}, 
see lowest panel in Fig.~\ref{fig:FFs}. The simulations are performed  in the center of mass of the initial fissioning nucleus. 
The dominant assumption so far in literature was that TKE can be evaluated at any 
separation between FFs after scission, using the procedure described here,  
and also that their intrinsic energies have well defined values determined at the 
scission configuration. The results of these first these study of the FFs properties within a framework free of any assumptions 
or approximations clearly demonstrate the invalidity  of these assumptions perpetuated in literature for many decades now. 
In particular, in Refs.~\cite{Bulgac:2019b,Bulgac:2020} it was conclusively demonstrated that the FFs deformation
properties evolve after scission too. 

In this paper I describe a simplified model 
of this excitation energy exchange between the FFs, which even though it is not aimed to be very accurate, it does 
it illustrates this new mechanism of the excitation energy sharing mechanism between FFs.

 \section{Coulomb interaction of fission fragments beyond scission}\label{sec:II}

This is a simple classical model of the dynamics of the FFs beyond scission, assumed as  
incompressible neutron and proton fluids with a Coulomb interaction between the two FFs.
\begin{align}
 H= &\frac{m(Z_1\dot{{\bm x}}_1^2+N_1\dot{{\bm y}}_1^2+Z_2\dot{{\bm x}}_2^2+N_2\dot{{\bm y}}_2^2)}{2} + \nonumber \\
& \frac{k_1|{\bm x}_1-{\bm y}_1|^2 + k_2 |{\bm x}_2-{\bm y}_2|^2}{2} + \frac{ e^2 Z_1 Z_2 }{ |{\bm x}_1-{\bm x}_2| },\label{eq:H}
\end{align}
where $m$ is the nucleon mass and $e$ the proton charge.
Here ${\bm x}_{1,2}$ and ${\bm y}_{1,2}$ are the proton and neutron center 
of mass coordinates, and $Z_{1,2}$ and $N_{1,2}$  are the proton and neutron numbers 
of the two FFs repectively. 
The two incompressible and frozen liquids in each fragment
can move with respect to each other in this model~\cite{Goldhaber:1948,Steinwedel:1950,Myers:1977,Ring:2004} 
with harmonic restoring forces.
There is no reason to further complicate unnecessarily this simple 
model, since accurate results including all possible other effects are 
already available~\cite{Bulgac:2016,Bulgac:2019b,Bulgac:2020}
and many more will follow~\cite{IA:2020}.
 
By introducing the coordinates
\begin{align}
&{\bm \xi}_1 = {\bm x}_1-{\bm y}_1,\quad {\bm \xi}_2={\bm x}_2-{\bm y}_2, \\
&{\bm \eta} = \frac{Z_1{\bm x}_1+N_1{\bm y}_1}{A_1}-\frac{Z_2{\bm x}_2+N_2{\bm y}_2}{A_2},\\
&{\bm \zeta} =\frac{Z_1{\bm x}_1+N_1{\bm y}_1+Z_2{\bm x}_2+N_2{\bm y}_2}{A} 
\end{align}
where $A_{1,2}=Z_{1,2}+N_{1,2}$ and $A=A_1+A_2$ the Hamiltonian becomes
\begin{align} 
H  =&\frac{\mu_1\dot {\bm \xi}^2_1}{2} + \frac{\mu_2\dot{\bm \xi}_2^2}{2}
+\frac{ \mu\dot{\bm \eta}^2}{2} +\frac{Am\dot{\bm \zeta}^2}{2} + \nonumber\\ 
 &\frac{k_1{\bm \xi}_1^2}{2} +\frac{k_2{\bm \xi}_2^2}{2} +\frac{e^2Z_1Z_2}{|{\bm \eta}+{\bm \xi}|},
\end{align}
where
\begin{align} 
&\mu_{k}=m\frac{Z_kN_k}{A_k}, \quad k=1,2, \quad \text{and} \quad \mu = m\frac{A_1A_2}{A}, \label{eq:mu}\\
&{\bm x}_1-{\bm x}_2 \equiv {\bm \eta} +\frac{N_1}{A_1}{\bm \xi}_1-\frac{N_2}{A_2}{\bm \xi}_2 =  {\bm \eta} +{\bm \xi}. \label{eq:xiii}
\end{align}  
Then the equations of motion become
\begin{align}
\mu_1\ddot{{\bm \xi}}_1 &= -k_1{\bm \xi_1}  +\frac{e^2Z_1Z_2N_1}{A_1}  \frac{({\bm \eta}+{\bm \xi})}{|{\bm \eta}+{\bm \xi} |^3}\nonumber \\
                             &\approx -k_1{\bm \xi_1}  +\frac{e^2Z_1Z_2N_1}{A_1}  \frac{{\bm \eta}}{|{\bm \eta} |^3} , \label{eq:xi1}\\
\mu_2\ddot{{\bm \xi}}_2 &= -k_2{\bm \xi }_2 - \frac{e^2Z_1Z_2N_2}{A_2} \frac{({\bm \eta}+{\bm \xi})}{|{\bm \eta}+{\bm \xi} |^3}\nonumber \\
                                   &\approx -k_2{\bm \xi }_2 - \frac{e^2Z_1Z_2N_2}{A_2} \frac{{\bm \eta}}{|{\bm \eta} |^3} ,\label{eq:xi2}\\
\mu  \ddot{{\bm \eta}}   &= \frac{e^2 Z_1 Z_2 ({\bm \eta}+{\bm \xi})}{ |{\bm \eta}+{\bm \xi} |^3}
                 \approx\frac{e^2 Z_1 Z_2 {\bm \eta}}{ |{\bm \eta} |^3},\label{eq:eta}
\end{align}
and where ${\bm \xi}$ has been defined in Eq.~\eqref{eq:xiii}.
As the center of mass coordinate ${\bm \zeta}$ is  not affected by interaction the corresponding equation can be ignored.  
Since $|{\eta }|\approx |{\bm x}_1-{\bm x}_2| \gg |{\bm \xi}_{1,2}|$ one can ignore ${\bm \xi}$ on the right hand sides of these equations and 
then these equations can be solved by quadrature. Notice the driving Coulomb force in Eqs.~(\ref{eq:xi1}-\ref{eq:xi2}) 
acts in opposite directions for ${\bm \xi}_{1,2}$. 

Assuming for simplicity that $k_{1,2}=\mu_{1,2}\omega^2$ 
the solutions are 
\begin{align}
&{\bm \eta}(\tau)= (0,0,R)(\cosh \tau+1), \label{eq:R_t}\\
 &t(\tau)=\sqrt{\frac{mR^3}{e^2Z_1Z_2}}(\sinh\tau+\tau),\label{eq:t_t} \\
&{\bm \xi}_k(t) = {\bm \xi}_k(0)\cos(\omega t)  \\
& +\int_0^tdt_1 \frac{\sin[\omega(t-t_1)]}{\omega}\frac{ C_k{\bm \eta}(t_1) }{ |{\bm \eta}(t_1)|^3 },
\quad C_k = \frac{e^2Z_1Z_2N_k}{A_k\mu_k}, \nonumber
\end{align}
where $2R=R_1+R_2$ is the distance between two touching spheres.
In this case ${\bm \xi}_k(0)\ne 0$, as the two FFs just before the neck is ruptured
can polarize each other, while they are practically at rest  $\dot{{\bm \xi}}(0)=0$, see also Fig.~\ref{fig:FFs}, as suggested
by the overdamped  character of the collective motion before neck rupture~\cite{Bulgac:2016,Bulgac:2019b,Bulgac:2020}. 
The initial polarization of the two FFs is given by the condition that the 
Coulomb force is balanced by the restoring force of the dipole modes.

\begin{figure}
\includegraphics[width=1\columnwidth]{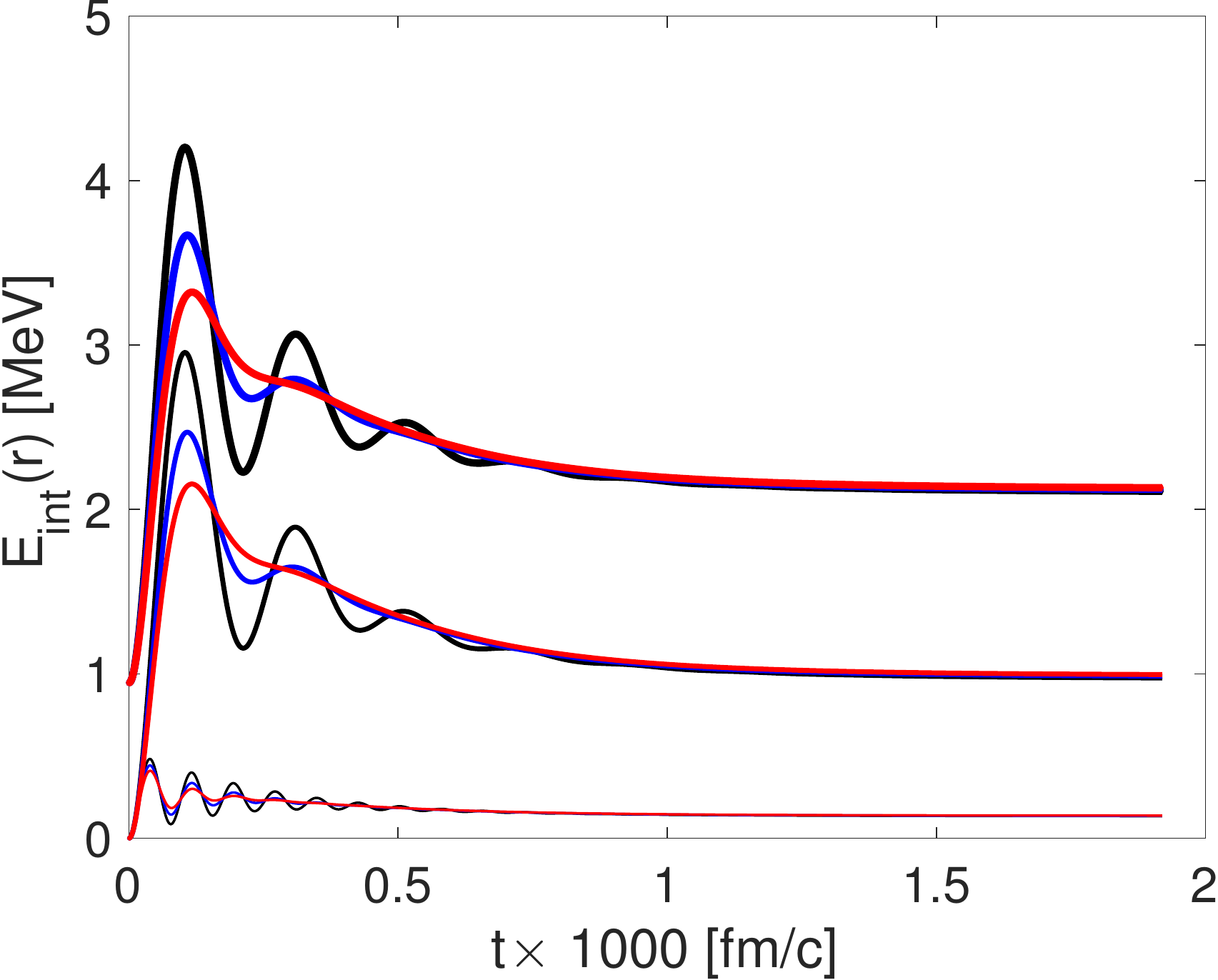}
\includegraphics[width=1\columnwidth]{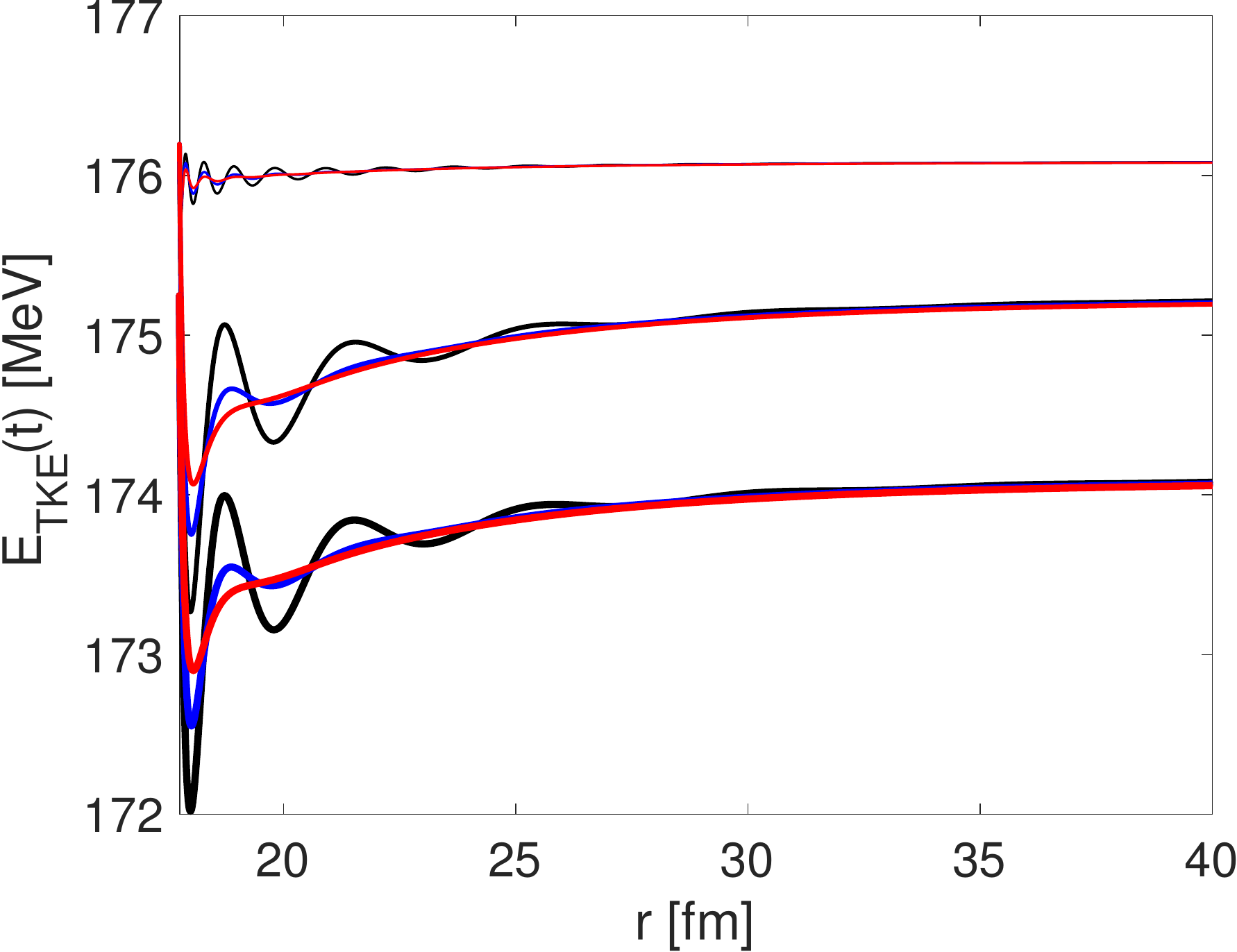}
\caption{ \label{fig:P} 
(Color online) An evaluation 
in the lowest order perturbation theory of FFs intrinsic energy $E_\text{int}(t)$ in the case of  $^{240}$Pu 
induced fission as a function of the separation between the FFs and also as a function 
of the time after scission.The initial separation and fragment charges were chosen so that 
$e^2Z_1Z_2/\eta$=176.2 MeV, with $Z_h=53$ and $Z_L=41$, typical for average FFs charges.
The initial value of the Coulomb energy is compatible with very large elongations of the FFs at scission.
The thinest and medium thickness lines correspond to the case ${\bm \xi}_{1,2} (0) = 0$
when  the FFs fragments are initially charge unpolarized,  
and $\hbar \omega_k=\sqrt{k_k/\mu_k}=16$ MeV and 6 MeV respectively.
The thickest lines   correspond to ${\bm \xi}_k(0) \ne 0$, thus to initially charge polarized  FFs
at scission, and $\hbar \omega_k=6$ MeV. The black, blue, and red lines correspond to damping 
$\hbar\gamma_1=\hbar\gamma_2 =(2, 4, 6)$ 
MeV respectively, see Eq.~(\ref{eq:xi12}-\ref{eq:xi22}).  
}
\end{figure}

Assuming that initial velocities are $\dot{{\bm \xi}}_{1,2}(0)=\dot{\bm \eta}(0)=0$, and $|{\bm \eta}(0)+{\bm \xi}(0)|=2R$,
one can define the total,  the intrinsic, the final kinetic energy of the fragments,  
and the Coulomb interaction between the fragments energies and for all times after scission $t>0$ 
\begin{align}
E_\text{tot}&= \left . \frac{ k_1{\bm \xi}^2_1+ k_2{\bm \xi}^2_2}{2}\right | _{t=0} +\frac{e^2Z_1Z_2}{2R}\nonumber \\
&=E_\text{int}(t)+E_\text{TKE}(t) >0,\label{eq:et}\\
E_\text{int}(t)&= \frac{  \mu_1\dot{\bm \xi}_1^2+ \mu_2 \dot{\bm \xi}_2^2 }{2}+\frac{k_1{\bm \xi}^2_1+ k_2{\bm \xi}^2_2}{2}> 0,\label{eq:int}\\
E_\text{TKE}(t)& = \frac{\mu \dot{{\bm \eta}}^2}{2}+\frac{e^2Z_1Z_2}{|{\bm \eta}+{\bm \xi}|}
 \rightarrow  \left . \frac{\mu \dot{{\bm \eta}}^2}{2} \right |_{t \rightarrow \infty}.  \label{eq:tke}
\end{align}
Here the intrinsic energy $E_\text{int}(t)$ stands for the combined additional excitation 
energy of both FFs acquired after 
scission, when the two fragments interact only through the long range Coulomb interaction. 
Thus the fragments end up excited and the (final) total kinetic energy of the fragments is less
than the initial Coulomb potential energy, see Fig.~\ref{fig:P}, as one would have naively expected. 
Eqs.~\eqref{eq:xi1},~\eqref{eq:xi2}, and~\eqref{eq:eta}) show that the energy exchange between intrinsic degrees of freedom 
${\bm \xi}_{1,2}$ and the relative fragment degrees of freedom ${\bm \eta}$ are 
controlled by Coulomb interaction alone. 

Since dipole oscillations are however damped it is appropriate to include this effect. The equations for the
intrinsic degrees of freedom ${\bm \xi}_k$ become in this case  
\begin{align}
\mu_1\ddot{{\bm \xi}}_1  \approx -k_1{\bm \xi_1}  - \mu_1\gamma_1\dot{\bm \xi}_1 + \frac{e^2Z_1Z_2N_1}{A_1}  \frac{{\bm \eta}}{|{\bm \eta} |^3} , \label{eq:xi12}\\
\mu_2\ddot{{\bm \xi}}_2  \approx -k_2{\bm \xi }_2 - \mu_2\gamma_2\dot{\bm \xi}_2 - \frac{e^2Z_1Z_2N_2}{A_2} \frac{{\bm \eta}}{|{\bm \eta} |^3} ,\label{eq:xi22}
 \end{align}
 At relatively small damping the main effect is the averaging of the oscillations with little change in the average 
value of the asymptotic value of $E_\text{TKE}(t)$, see Fig.~\ref{fig:P}, and  an 
additional increase of the internal excitation energy of the fragments, 
beyond that acquired during the descent from saddle-to-scission, of up to a few MeVs
in case of strong damping. In these simulations one observes that the light FF typically 
emerges very elongated, see Refs.~\cite{Bulgac:2019b,Bulgac:2020} and Fig.~\ref{fig:FFs},  
and in that case the energy of the dipole resonance can be 
rather low, similar to a pygmy resonance energy, and  in that case 
${E}_\text{int}$ increases noticeably when $\hbar\omega$ decreases, see Fig.~\ref{fig:P}.

One should also note that even though the dipole resonances are excited
in the FFs, most likely this is not going to lead to emission of relatively high-energy gamma rays, since this excitation
energy is dissipated in a time interval much shorter $ \approx 10^{-21}$ sec.  than the times required 
to emit a photon $\approx 10^{-14}\ldots10^{-3}$ sec. see Refs. ~\cite{Vandenbosch:1973,Wagemans:1991,Krane:1987} and 
particularly G\"onnenwein's lecture notes~ \cite{Gonnenwein:2014}.    Nevertheless,
experiments point to the observation ``that the intensity of the $\gamma$-ray energy above 
5 MeV is sensitive to the species of fissile nuclei"  and to the likely ``population of pygmy resonances"~\cite{Chyzh:2013}.

This model neglects the excitation of other 
collective modes. When FFs 
are accelerated, in their own non-inertial reference frame they experience a force, which tends 
to pile up the nuclear matter at the edges facing each other, similarly to a what happens to
an accelerated vessel with water. One thus expects that both iso-scalar and iso-vector 
modes are excited as seen in realistic 
simulations~\cite{Simenel:2014,Bulgac:2016,Bulgac:2019b,Bulgac:2020}. 
This model also neglects that the FFs large deformations at scission change
significantly after scission also~\cite{Bulgac:2019b,Bulgac:2020}. Since the shapes of the FFs evolve 
in time even after scission, the collective excitation energy stored in these modes is still 
dissipated due to the one-body dissipation mechanism~\cite{Blocki:1978}. 
The decay of the giant resonances into more complex  particle-hole excitations,  is typically
described by the spreading width $\Gamma^\downarrow$~\cite{Bertsch:1983}, 
which is a de-excitation mechanism somewhat independent of the one-body dissipation due to 
nuclear large amplitude collective motion of the FFs. 
Since the during the descent from the saddle to scission the motion is strongly overdamped,
close to the scission configuration the kinetic energy of the fragments in the fission direction is 
negligible~\cite{Bulgac:2019b,Bulgac:2020} and $E_\text{TKE}(0)\approx e^2Z_1Z_2/2R $. 
This is contrast with phenomenological calculations~\cite{Sierk:2017,Usang:2019}, when 
the FFs have a significant kinetic energy at scission. I am 
aware of a single instance where the dipole excitation of the FFs was examined earlier~\cite{Mustafa:1971}, 
where a relatively small increase of $E_\text{int}$ was found.

In most phenomenological models~\cite{Brosa:1990,Wagemans:1991,Sierk:2017,Usang:2019,Sadhukhan:2017} 
and in the time-dependent 
generator coordinate method~\cite{Regnier:2016,Regnier:2019}
the collective motion before scission is only partially, if ever, 
damped, and at scission the two FFs have 
a kinetic energy of the order of $10\ldots 15$ MeV,
with the exception of Smoluchowski approaches~\cite{Grange:1983,Weidenmuller:1984,
Randrup:2011,Randrup:2011a,Randrup:2013,Ward:2017,Albertsson:2020} and the unrestricted 
TDDFT framework~\cite{Bulgac:2016,Bulgac:2019b,Bulgac:2020}.  In the unrestricted TDDFT framework
the excitation of all collective modes  by the Coulomb interaction between FFs  and 
a significant amount of their damping mechanism after scission are accounted for.  
One might also consider the case when the isovector mass is different from the bare mass. That 
would require a simple replacement $m\rightarrow m^*$ in the definition of reduced masses $\mu_{1,2}$, see Eq.~\eqref{eq:mu}. \\

\section{Conclusion}

While the model presented here is simplified and classical, it is pretty realistic. 
It is straightforward to implement into such a model
various  deformations. 
At the same time it is unnecessary to perform such involved model calculations when 
realistic calculations are available~\cite{Bulgac:2016,Bulgac:2019,Bulgac:2019b,Bulgac:2020} 
and new ones are in the pipeline. The only 
relevant question is that of the correct interpretation of those realistic results, for which a 
simple model is particularly useful. I have shown 
here that the FFs exchange up to several MeVs of
excitation energy, after they ceased to exchange nucleons, and up to
relatively large separations, due to the long range character of the Coulomb interaction between them. 
This excitation energy mechanism leads to slightly smaller final TKE of the FFs. 
Similar effects are expected in the case of fragments emerging 
in heavy-ion reactions.

\vspace{0.4cm}

\section*{\bf Acknowledgements} 
I thank I. Abdurrahman for preparing Fig.~\ref{fig:FFs}. 
I also thank G.F. Bertsch, I. Stetcu, and N. Carjan for discussions.
This work was supported by U.S. Department of Energy,
Office of Science, Grant No. DE-FG02-97ER41014 and in part by NNSA
cooperative Agreement DE-NA0003841. 


\providecommand{\selectlanguage}[1]{}
\renewcommand{\selectlanguage}[1]{}

\bibliography{latest_fission}

\end{document}